\documentstyle[amssymb,aps,multicol,epsf]{revtex}

\begin{document}
\draft

\title{
Modern architecture of random graphs: Constructions and correlations
}

\author{
S.N. Dorogovtsev,$^{1, 2, \ast
}$ J.F.F. Mendes,$^{1,\dagger}$ and A.N. Samukhin$^{1, 2, \ddagger
}$
}

\address{
$^{1}$ Departamento de F\'\i sica and Centro de F\'\i sica do Porto, Faculdade 
de Ci\^encias, 
Universidade do Porto\\
Rua do Campo Alegre 687, 4169-007 Porto, Portugal\\
$^{2}$ A.F. Ioffe Physico-Technical Institute, 194021 St. Petersburg, Russia 
}

\maketitle

\begin{abstract}
1. Basic constructions. 
2. Equilibrium and nonequilibrium networks. 
3. Equilibrium uncorrelated networks. 
4. Nonequilibrium nongrowing scale-free nets. 
5. Types of correlations. 
6. When pair correlations are important. 
7. When loops are important. 
8. Pair degree-degree correlations in growing networks. 
9. How to construct an equilibrium net with given degree-degree correlations. 
10. How to construct a growing scale-free net with a given clustering (towards a real-space renormalization group for networks)\vspace{10pt}.   
      
\end{abstract}


\centerline{\em (a talk given at XVIII Sitges Conference)}

\begin{multicols}{2}

\narrowtext



\section{Basic constructions}\label{s-basic} 

The basis of network science are construction procedures. 

There are four main constructions of networks (and a large number of their versions and variations): 

\begin{itemize} 

\item[(1)] 
{\em Classical random graphs} of Erd\"{o}s and R\'enyi \cite{er59} ($\sim 1960$). 
\\
\hspace*{\parindent} \ \, 
In general terms, these are 
graphs with a fixed number of vertices, which are connected, at random, by edges. Classical random graphs have a Poisson degree distribution. 

\item[(2)] {\em The random graphs with a given degree distribution}, or, more rigorously, the labeled random graphs with a given degree sequence, which are also called {\em the configuration model}, see a long list of 
references \cite{mr95} starting from 1972 (!). 
\\
\hspace*{\parindent} \ \, 
In general terms, these are graphs, maximum random under the constraint that their degree distribution is a given function. 
When this function is Poisson, (2) is reduced to (1).   

\item[(3)] {\em Small-world networks} which were introduced by Watts and Strogatz \cite{ws98} (1998). 
\\ 
\hspace*{\parindent} \ \, 
In general terms, this is the superposition of regular lattices and the classical random graphs. The small-world networks combine the compactness of the classical random graphs and the strong clustering of many regular lattices. 
This construction realize a simple way of the introduction of the particular type of correlations (the clustering) in networks.   


\item[(4)] {\em Networks, growing under the mechanism of preferential linking} 
which were introduced by Barab\'asi and Albert \cite{ba99} (1999). 
\\ 
\hspace*{\parindent} \ \, 
Into this class of constructions, we include much more general models of networks, evolving by the addition of vertices and edges, than the original Barab\'asi-Albert model. For example, random linking is a particular case of the preferential linking.   

\end{itemize}

\section{Equilibrium and nonequilibrium networks}\label{s-equil} 

As the objects of statistical physics, networks can \vspace{6pt} be 
\\ 
\centerline{{\em equilibrium} (1, 2, 3) 
$\longleftrightarrow$ {\em nonequilibrium} (4).}\vspace{8pt}$\phantom{.}$ 

The second division is according to the form of degree distributions\vspace{6pt}: 
\\ 
\centerline{rapidly decreasing degree distributions} 
\\
\centerline{$\longleftrightarrow$}  
\\ 
\centerline{{\em fat-tailed degree distributions}}
\\
\centerline{(with divergent higher moments).}\vspace{8pt}$\phantom{.}$  

The third division is \vspace{6pt} into 
\\ 
\centerline{{\em uncorrelated} (1, 2) $\longleftrightarrow$ {\em correlated} (3, 4) networks.}\vspace{8pt}$\phantom{.}$   

\begin{figure}
\epsfxsize=65mm
\centerline{\epsffile{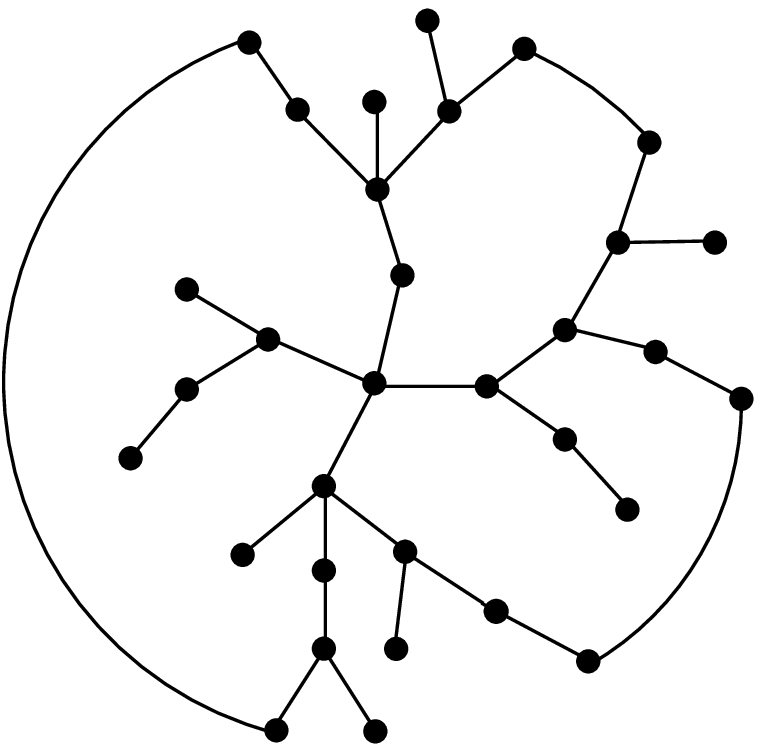}}
\caption{
}
\label{f1}
\end{figure}

Uncorrelated random graphs with a given degree distribution (2) have a tree-like (locally) structure. The structure of these networks, from the point of view of a vertex, belonging to a ``giant connected component'', is schematically shown in Fig. \ref{f1}. Notice that loops are present only at a large scale. 


\section{Equilibrium uncorrelated networks}\label{s-eq}

\subsection{Configuration construction} 

In statistical \vspace{8pt} mechanics 

\centerline{
{\em a random network is a statistical ensemble of graphs}\vspace{8pt}. 
}

\noindent
That is, what is observed are members of statistical ensembles, or, which is the same, their particular realizations. 

A random graph with a given degree distribution (the configuration model (2)) is a statistical ensemble, which includes all possible networks with a given degree distribution and given numbers of vertices $N$ and edges $L$. Each member of the ensemble is taken with an equal statistical weight. 

In fact, these are maximum random graphs under the constraint that their degree distributions are given functions. 

\begin{figure}
\epsfxsize=48mm
\centerline{\epsffile{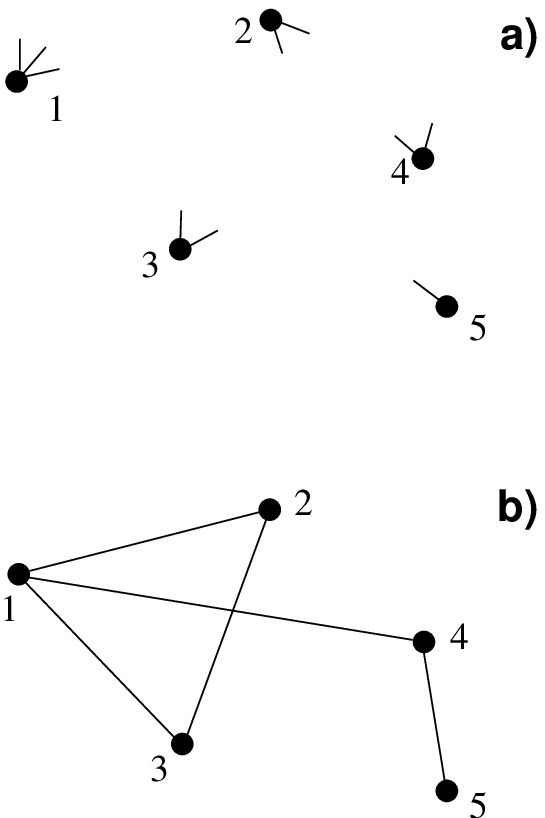}}
\caption{
}
\label{f2}
\end{figure}

To construct a large typical realization of this graph, one can use the following procedure (see Fig. \ref{f2}): 

\begin{itemize} 

\item[(i)] 
Prepare $N$ vertices, $i=1, \ldots , N$. 

\item[(ii)] 
Attach to each vertex $k_i$ ends of edges taken from the given distribution $P(k)$. 

\item[(iii)] 
Connect at random ends of pairs of quills of the resulting ``family of hedgehogs''.

\end{itemize}  


\subsection{Dynamical construction}

These graphs can be also constructed dynamically by many ways 
(see Refs. \cite{bck01,dms02c,bb02b}). 
For example, let $N$ and $L$ be fixed.  

\begin{itemize} 

\item
At each time step, rewire a randomly chosen end of a randomly chosen edge to a preferentially chosen vertex: the probability that the edge end becomes attached to a vertex of degree $k$ is proportional to a function $f(k)$ of this degree. 

\end{itemize}  

\noindent
In the long-time limit, the network will have a stationary degree distribution which is determined by the form of $f(k)$.  

For example, if $f(k) \propto k+1-\gamma $ for $k \to \infty$, 
\\
the network is scale-free at some critical value of the mean degree, $\overline{k}_c$, $P_c(k) \propto k^{-\gamma}$ (see Fig. \ref{f3}). 
\\
For smaller $\overline{k}$, this power law has a size-independent cutoff. 
\\ 
For $\overline{k}>\overline{k}_c$, a finite fraction of edges is condensed on a single (or a few) vertices, and the rest vertices have the same degree distribution as at the critical point \cite{bck01,dms02c,bjj02}.


\begin{figure}
\centerline{\begin{picture}(100,150)
\put(38,140){$\overline{k}$} 
\put(-19,118){CONDENSED \ \  PHASE} 
\put(56.5,100){$P_c(k)$ \& $N(\overline{k}-\overline{k}_c)$} 
\put(56.5,90){condensed} 
\put(56.5,80){on a ``single'' vertex}
\put(-20,64){$\overline{k}=\langle f(k) \rangle$\,:} 
\put(38.5,64){$\overline{k}_c$ \ \ \ $P_c(k) \propto k^{-\gamma}$} 
\put(-20,60){\line(2,0){125}}
\put(-2,42.5){NORMAL \ \ \ PHASE} 
\put(56.5,30){exponential} 
\put(56.5,20){size-independent} 
\put(56.5,10){cutoff of $P_c(k)$} 
\put(38.2,0){$0$\ $-$} 
\put(50,0){\vector(0,2){150}}
\end{picture}}
\caption{
} 
\label{f3}
\end{figure}


\section{Nonequilibrium nongrowing scale-free nets}\label{s-types}

\begin{figure}
\epsfxsize=79mm
\centerline{\epsffile{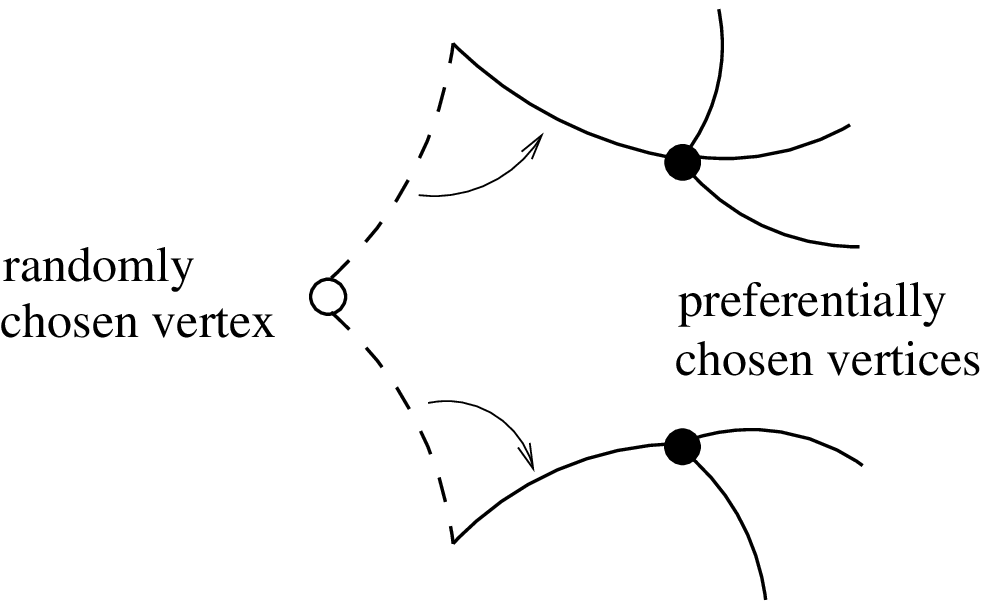}}
\caption{
}
\label{f4}
\end{figure}

{\em Even networks with a constant number of vertices and edges may be nonequilibrium and may be ``scale-free''}. See the following illustrative example of a network with finite numbers of vertices $N$ and edges $L$, $\overline{k}=2L/N$ (Fig. \ref{f4}).  

\begin{itemize} 

\item[(i)] 
At each time step choose, at random, a vertex. 

\item[(ii)]  
Rewire all the edge ends from this vertex to preferentially chosen vertices.

\item[(iii)] 
For each of these ends, the probability that it becomes attached to a vertex of degree $k$ is proportional to $k+A$. 

\end{itemize} 

The resulting network (in the long-time limit) is {\em scale free}: $P(k) \propto k^{-\gamma}$ with exponent $\gamma=2+A/\overline{k}$.


Another example is: 

\begin{itemize} 

\item[(i)] 
At each time step choose, at random, a vertex. 

\item[(ii)]  
Rewire all the edge ends from this vertex to vertices, which are chosen according to the following rule: 
\\
(a) with probability $p$, an edge end is rewired to a randomly chosen vertex, 
\\
(b) with probability $1-p$, an edge end is rewired to a vertex chosen with preference, proportional to its degree $k$. 

\end{itemize} 

The resulting network (in the long-time limit) is {\em scale free}: $P(k) \propto k^{-\gamma}$ with exponent $\gamma=2+p/(1-p)$.

\section{Types of correlations}\label{s-types} 

The very particular kinds of correlations are: 

\begin{itemize} 

\item[(a)] 
{\em Correlations of the degrees of nearest-neighbor vertices}. This means that the joint degree-degree distribution of end vertices of edges does not factor: 

\begin{equation}
P(k,k') \neq \frac{kP(k)k'P(k')}{[\sum kP(k)]^2}
\, .   
\label{1}
\end{equation} 

\item[(b)] 
{\em Loops}, and so, in particular, {\em clustering}, since the clustering, in general terms, is the concentration of loops of length three in a network. 

\end{itemize} 

In uncorrelated networks, loops and clustering are only size effects\vspace{8pt}. 

{\em The growth of a network produces a wide spectrum of correlations which cannot be reduced to only} (a) {\em and} (b).

\section{When pair correlations are important}\label{s-corr_imp} 

For example,

\begin{itemize} 

\item
In uncorrelated scale-free equilibrium networks with $\gamma \leq 3$, an average shortest-path length $\overline{\ell}(N)$ grows with a number of vertices $N$ slower than $N$ (a super-small-world effect \cite{ch02}). 

\item[] \centerline{$\longleftrightarrow$}

\item
\vspace{-24pt}On the other hand, as a rule, pair correlations in real nets are such that the average degree of the nearest neighbors of a vertex is a decreasing function 
$\overline{k}_{nn}(k)$ of its degree $k$. In such a situation, a standard relation $\overline{\ell} \sim \ln N/\ln\overline{k}$ is valid. 

\end{itemize} 

In a number of real nets, $\overline{\ell}(N)$ is nearly constant in a wide range of variation of $N$ (see, e.g., networks of metabolic reactions 
\cite{jmbo01}). The reason for this constancy is not ``the super-small-world effect'' but the increase of $\overline{k}(N)$ with $N$: these nets become more ``dense'' as they grow. 

Notice that in all the real scale-free nets, the ratio 
$\overline{\ell}_{real}/(\ln N/\ln\overline{k})$ is in a surprisingly narrow range between $1/2$ and $2$.

\section{When loops are important}\label{s-loops_imp}

In statistical physics, and so in networks, loops are responsible for fluctuation phenomena in a critical region.

\section{Pair degree-degree correlations in growing networks}\label{s-pair} 

{\em Pair degree-degree correlations is the generic property of growing networks.} 

For example, consider a scale-free citation graph. By definition, a citation graph is a growing network, in which new connections between already existing vertices are impossible: new edges connect new and old vertices. 

\begin{figure}
\epsfxsize=72mm
\centerline{\epsffile{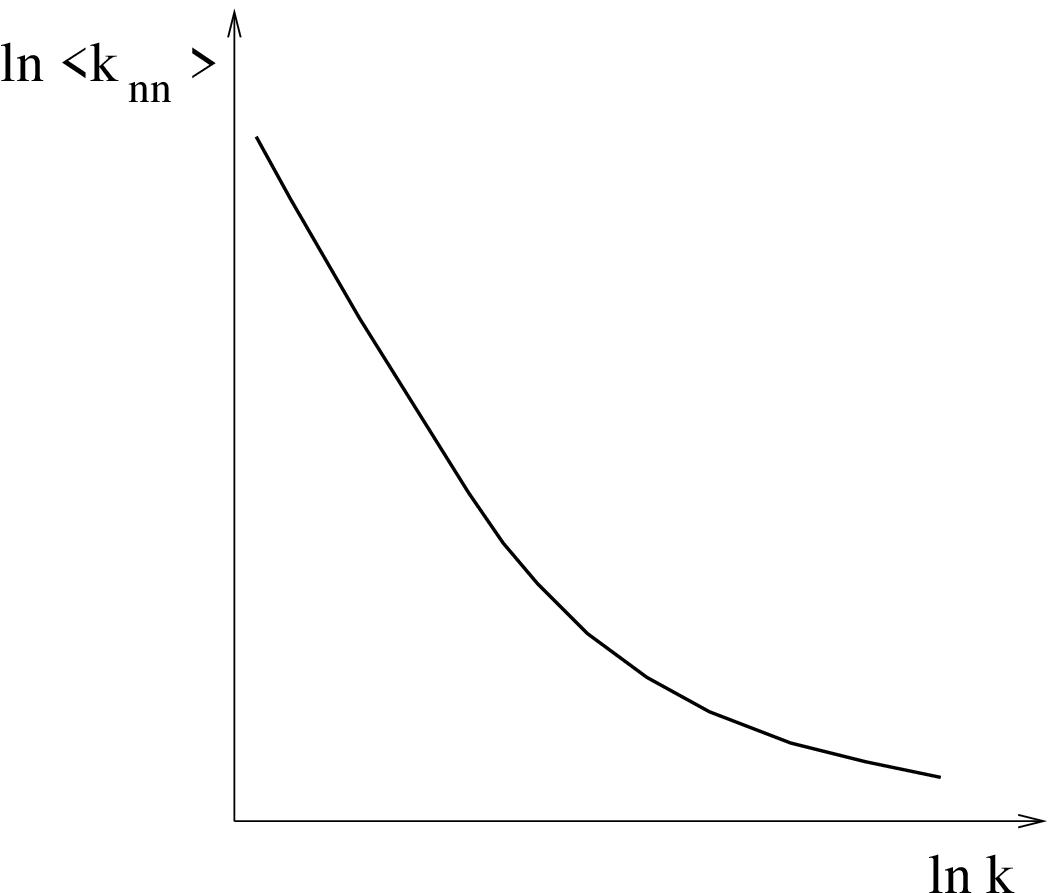}}
\caption{
}
\label{f5}
\end{figure}

Let $\gamma<3$. Then, the asymptotic expression for the joint degree-degree distribution \cite{kr00c} for nearest neighbor vertices is   

\begin{equation}
P(k,k') \stackrel{k \gg k' \gg 1}{\propto} k^{-(\gamma-1)} k'^{-2}
\, .  
\label{3}
\end{equation} 
This gives the following form of the average degree of the nearest neighbors of a vertex as a function 
$\overline{k}_{nn}(k)$ of its degree $k$ (see Fig. \ref{f5}): 

\begin{itemize}

\item
at small enough $k$, $\overline{k}_{nn}(k) \propto k^{-(3-\gamma)}$; 

\item
at larger $k$, $\overline{k}_{nn}$ logarithmically depends on $k$.  

\end{itemize}


So, in this respect, the case $\gamma=3$, that is, the Barab\'asi-Albert model, is exceptional: in this event, the pair correlations are small. 

When $\gamma$ of a citation graph is greater than $3$, the slope of the curve in Fig. \ref{f5} becomes positive. Even if the attaching is random (the net is exponential) the strong pair correlations are present: for the highest degree $k$, 
$\overline{k}_{nn}(k) \sim k$, and $\overline{k}_{nn}(1)= \overline{k}$. 

One can easily change pair correlations by changing, e.g., an additional attractiveness $A$ in the preference function $k+A$. One can also regulate correlations 
by introducing, in addition, a new channel of linking of already existing edges. So, the slope of the dependence $\overline{k}_{nn}(k)$ strongly depends on a model. 

One may say, the smallness of pair correlations in growing networks is a rare event and not a standard situation. The standard mechanisms of the network growth necessarily produce these correlations.

\section{How to construct an equilibrium net with given degree-degree correlations}\label{s-equil_given}

\subsection{Configuration construction}\label{ss-conf}

In the spirit of the configuration model, one can construct the equilibrium net with a given pair degree-degree $P(k,k')$ and a given number $N$ of vertices as a statistical ensemble. Members of this statistical ensemble are all networks with a given $P(k,k')$. Each member is taken with an equal statistical weight. 

These are networks, maximum random under the constraint that their pair degree-degree distribution is equal to a given $P(k,k')$ \cite{n02d}.

To construct a large typical realization of this graph for large $N$, one can use the following way. 

Notice that in these networks, the ordinary degree distribution follows from the degree-degree distribution: 

\begin{equation} 
\sum_k P(k,k^\prime) = \frac{kP(k)}{\sum_k kP(k)} 
\, .  
\label{4}
\end{equation} 
This relation gives the function $P(k)$ up to a constant factor, 
which, in turn, can be obtained from the normalization condition $\sum_k P(k) = 1$. Thus we obtain  
\begin{equation} 
\overline{k} = \left[\sum_{k,k^\prime}P(k,k^\prime)/k\right]^{-1}
\, ,  
\label{5}
\end{equation} 
and so the total number of edges is known, $L=\overline{k}N/2$. 

So, we have $N$ and $L$ fixed and can formulate the construction procedure as it follows: 

\begin{itemize}

\item[(i)] Using the given $P(k,k^\prime)$ find the total number of edges, $L$. 

\item[(ii)] Create the infinite number $L$ of pairs of integers $(k,k^\prime)$ distributed as $P(k,k^\prime)$. These pairs play the role of $L$ edges with ends of degrees $k$ and $k^\prime$. 

\item[(iii)] Select, at random, groups of ends of degree $k$, 
each one consisting of $k$ ends. 
Tie them in bunches, each of $k$ tails. These bunches are vertices of degree $k$. 

\end{itemize}
One can check that this procedures is possible in the thermodynamic limit, and it generates tree-like (locally) graphs with the low number of loops like in uncorrelated networks (see Fig. \ref{f1}).

\subsection{Dynamical construction}\label{ss-dynamical} 

An equilibrium random graph with pair degree-degree correlations of nearest neighbor vertices also may be constructed dynamically. 

Let us introduce a new category, {\em pair preference} and {\em a pair preference function}, $f(k,k')$. 
This is a more general notion than a one-vertex preference, where the probability that the edge becomes attached, $f(k)$, depends only on the degree of a vertex. Indeed, an edge has not one but two ends. In general, they must become attached not to one but to two vertices, and so it is a combination of two degrees that is natural form of preference. 

Let $N$ and $L$ be fixed. 

\begin{figure}
\epsfxsize=78mm
\centerline{\epsffile{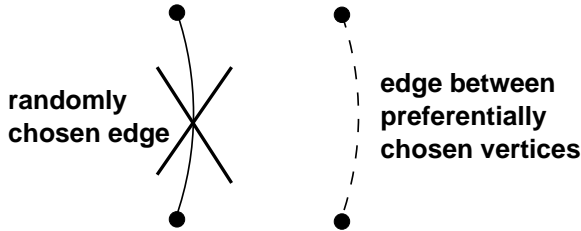}}
\caption{ 
A randomly chosen edge become removed to a pair of vertices of degrees $k$ and $k'$, which is chosen with probability, proportional to a preference function $f(k,k')$. 
}
\label{f6}
\end{figure}

\begin{itemize} 

\item 
At each time step, we choose, at random, an edge and remove it to a pair of vertices of degrees $k$ and $k'$, which is chosen with probability, proportional to a given function $f(k,k')$ (see Fig. \ref{f6}).  

\end{itemize} 

At long times, the network approaches a state with a stationary pair degree-degree degree distribution of nearest neighbor vertices, $P(k,k')$, which is determined by the form of $f(k,k')$. 

(For growing networks, a similar procedure was discussed in the talk of L. Pietronero at this conference.) For a related construction procedure, see Ref. \cite{bl02c}.


\section{How to construct a growing scale-free net with a given clustering 
\\ 
(towards a real-space renormalization group for networks)}\label{s-growing_given} 

Here we present a simple way to construct random growing correlated clustered, scale-free, etc. networks. The procedure is based on the stochastic transformation of each edge of a network at each step of the evolution. 

\begin{figure}
\epsfxsize=55mm
\centerline{\epsffile{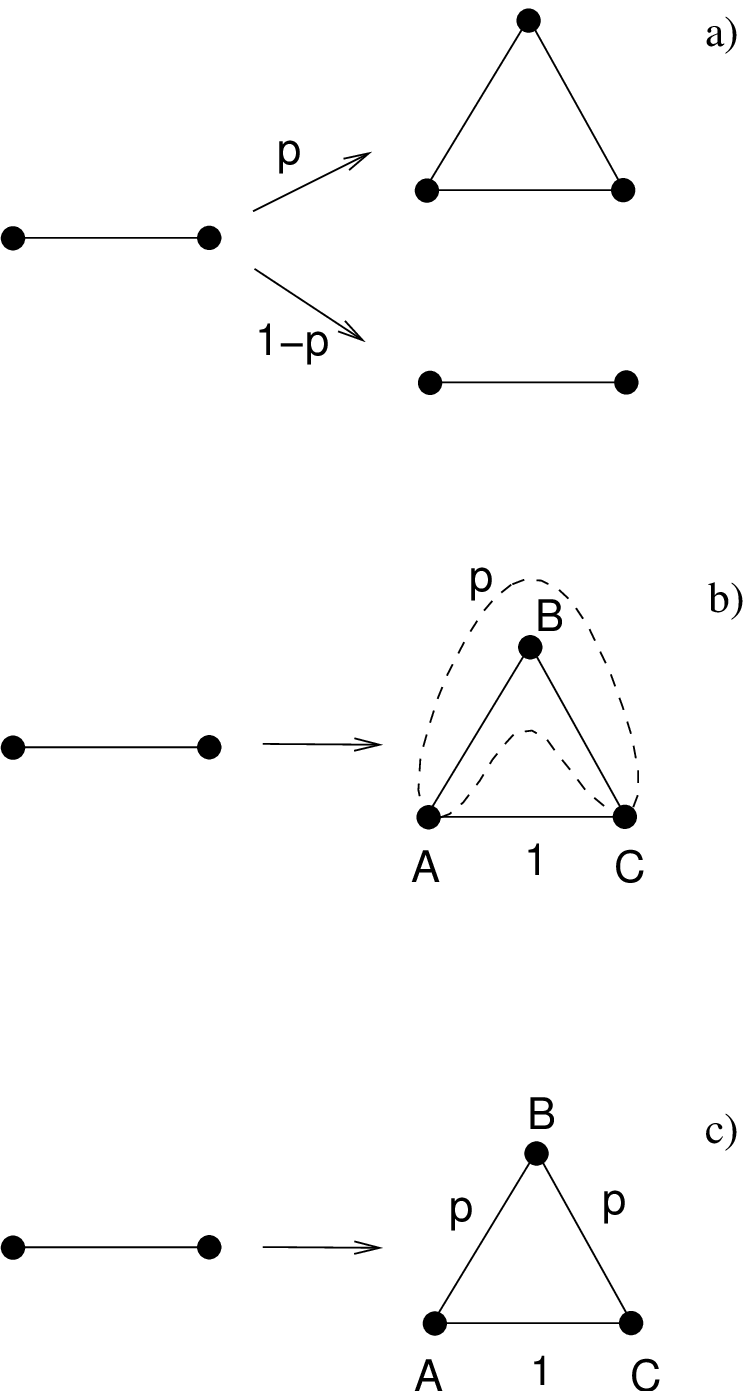}}
\caption{ 
}
\label{f7}
\end{figure}

Numerous versions of this process are possible. 
A simple particular case is shown in Fig. \ref{f7} (a). With some probabilities, a vertex is transformed to distinct configurations of edges and vertices.


If $p=1$, we have a deterministic growing graph \cite{dgm02}. 
Another representation of Fig. \ref{f7} (a) is given by Fig. \ref{f7} (b). 
For $p \to 0$ but $p \neq 0$, this construction coincide with a known 
random network, growing by attaching of new vertices to randomly chosen edges. 
Notice that the clustering coefficients of these nets are finite in the large network limit.  

Another possible construction, shown in Fig. \ref{f7} (c), coincides with the Barab\'asi-Albert model in the case $p \to 0$ but $p \neq 0$. 
 
\begin{figure}
\epsfxsize=75mm
\centerline{\epsffile{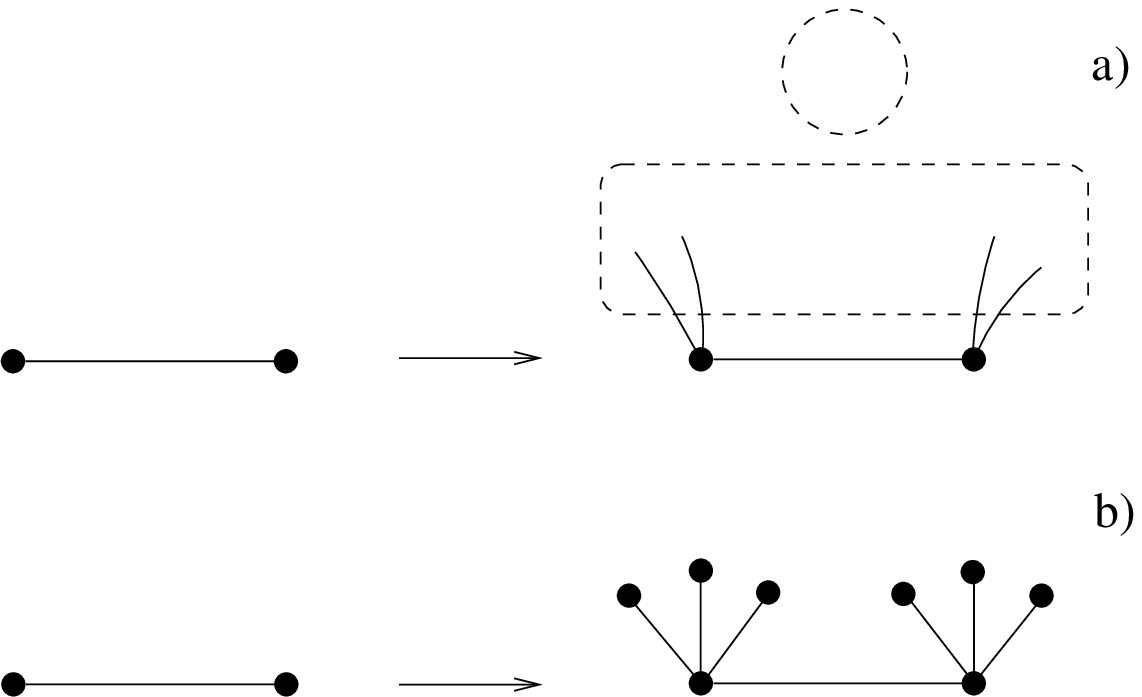}}
\caption{ 
}
\label{f8}
\end{figure}

We show only particular configurations on the left-hand parts of 
Fig. \ref{f7}. In general, they can be of a very general form (see Fig. \ref{f8} (a)), and so, they can be even trees (Fig. \ref{f8} (b)) \cite{jkk02}.  
Notice, that we used only the transformations that produce {\em networks with the small-world effect and fat-tailed degree distributions} (non-fractals: $\overline{\ell}(N) \propto \ln N$ grows slower than any power of $N$). 
Compare them, e.g., with contrasting fractal trees \cite{va97}. 
 

In fact, the presented transformations are very similar to those which are used in a standard real space renormalization group for 
disordered lattice models. This is also called ``positional renormalization group''. The technique of the real space renormalization group is well developed for disordered lattices. So, its application to these growing random networks is a standard routine\vspace{8pt}. 

Note that all the constructions in this talk are analytically treatable\vspace{10pt}. 

Thus,   
{\em correlations in networks are natural}.   
\\ 

\noindent
$^{\ast}$      Electronic address: sdorogov@fc.up.pt\\
$^{\dagger}$   Electronic address: jfmendes@fc.up.pt\\
$^{\ddagger}$  Electronic address: samukhin@fc.up.pt

\end{multicols}

\end{document}